\documentclass[
amsmath,
reqno,tbtags,
psamsfonts,
reprint,showpacs, 
prl, superscriptaddress]{revtex4-1}

\usepackage[english]{babel}
\usepackage{stmaryrd}
\usepackage{graphicx}
\usepackage{upgreek}
\usepackage{dcolumn}
\usepackage{bm}
\usepackage[dvipsnames]{xcolor}
\usepackage[capitalise]{cleveref}

\newcommand\eps{\ensuremath{\varepsilon}}
\newcommand{\msd}{\ensuremath{\delta r^2}}
\newcommand{\av}[1]{{\ensuremath{\left\langle #1 \right\rangle}}}

\newcommand{\fluid}{\ensuremath{_{F}}}
\newcommand{\matrx}{\ensuremath{_{M}}}
\newcommand{\FF}{\ensuremath{_{FF}}}
\newcommand{\MF}{\ensuremath{_{MF}}}

\newcommand{\dd}{\mathrm{d}}

\newcommand{\reduced}{\ensuremath{^*}}

\newcommand{\crit}{\ensuremath{_{c}}}
\newcommand{\eff}{\ensuremath{_\text{eff}}}

\newcommand{\new}[1]{#1}

\usepackage{times}

\begin{document}

\title[]{Crowding of interacting fluid particles in porous media through
    molecular dynamics: \\breakdown of universality for soft interactions}
\bigskip
\author{Simon K. Schnyder}
\email{skschnyder@gmail.com}
\affiliation{Fukui Institute for Fundamental Chemistry, Kyoto University, Kyoto 606-8103, Japan}
\affiliation{Institut f\"{u}r Theoretische Physik II, Heinrich-Heine-Universit\"{a}t 
D\"{u}sseldorf, Universit\"{a}tsstra{\ss}e 1, 40225 D\"{u}sseldorf, Germany}

\author{J{\"u}rgen Horbach}
\email{horbach@thphy.uni-duesseldorf.de}
\affiliation{Institut f\"{u}r Theoretische Physik II, Heinrich-Heine-Universit\"{a}t 
D\"{u}sseldorf, Universit\"{a}tsstra{\ss}e 1, 40225 D\"{u}sseldorf, Germany}

\date{January 27, 2018}

%
\begin{abstract}
Molecular dynamics simulations of interacting soft disks confined in
a heterogeneous quenched matrix of soft obstacles show dynamics which is fundamentally
different from that of hard disks. The interactions
between the disks can enhance transport when their density is increased,
as disks cooperatively help each other over the finite energy barriers
in the matrix.  
The system exhibits a transition from a diffusive to a
localized state but the transition is strongly rounded. 
Effective exponents in
the mean-squared displacement can be observed over three decades in time
but depend on the density of the disks and do not correspond to asymptotic behavior in the vicinity of a critical point, 
thus showing that it is incorrect
to relate them to the critical exponents in the Lorentz model scenario. 
The soft interactions are therefore responsible for a breakdown of the universality of the dynamics.

\end{abstract}
%


\maketitle

The transport of matter in heterogeneous porous materials is widespread, e.g. crowding phenomena in biology \cite{Weiss2014, Hofling2013, Sokolov2012, Saxton2012,Berry2014,Sung2006}, ion-conduction in silicate glasses \cite{Bunde1998, Horbach2002}, hydrology \cite{Bijeljic2011, Kang2016}, and other situations \cite{Gleiter2000, Brenner1993, Benichou2010, Ben-Avraham2000, Konincks2017}. Such systems consist of at least two components, characterised by a strong separation of time scales. The more mobile component often exhibits anomalous diffusion, i.e. its mean-squared displacement $\msd(t)$ grows nonlinearly over long periods of time. 
Often, anomalous diffusion can be characterised by an effective exponent, $\msd(t)\sim t^x$ with typically $x<1$, with a wide range of values for $x$ found. 
It remains unclear if the observed exponents merely represent transient behavior or whether they can be connected to a universal behavior with a well-defined exponent. This question will be addressed here.

A paradigm for the modeling of transport in heterogeneous media is the Lorentz Model (LM) \cite{Lorentz1905,Beijeren1982,Hofling2006, Hofling2007,
Hofling2008, Bauer2010,Spanner2016,Jin}, where anomalous diffusion arises as a universal long-time limit: In its simplest version, a single mobile particle moves in the static void space formed by overlapping hard disk obstacles. At low obstacle density, the mobile particle freely explores the system and exhibits regular diffusion. At high densities, it becomes trapped in finite pockets of obstacles. In-between there is a localisation transition, where the void space of the system stops to percolate, the system becomes self-similar, and anomalous diffusion occurs. This transition is a dynamic critical phenomenon and the exponent of the anomalous diffusion 
is universal~\cite{Ben-Avraham2000,Hofling2013}.

The LM can be generalized by introducing interacting mobile particles and soft instead of hard interactions, making it  more comparable to realistic porous materials.
But how such generalizations change the dynamics is not well understood. 
An extension~\cite{Krakoviack2005, Krakoviack2007, Krakoviack2009, Krakoviack2011} of the mode coupling theory of the glass transition (MCT) \cite{Gotze2008} predicts the LM localization transition to persist for interacting fluids in porous media, i.e. that the nature of the transition is unchanged by 
the interactions, and that the critical behavior is qualitatively the same.
Evidence from simulations of model porous media with interacting particles both with hard and soft interactions has been inconclusive \cite{Kurzidim2009, Kurzidim2010, Kurzidim2011, Kim2009, Kim2010, Kim2011}. 
While a localization transition and extended anomalous diffusion are observed, the exponents seldomly match the predictions. Still, so far it seemed to be evident that porous media with soft or hard potentials are qualitatively equivalent, even though 
 energy barriers in systems with soft potentials are finite and therefore are  crossable by soft particles~\cite{Skinner2013,Schnyder2015,Schnyder2016}.

Here, we perform molecular dynamics simulations of interacting soft disks confined in a soft heterogeneous matrix of obstacles. By systematically moving away from the single-particle case, we investigate how interactions between the mobile particles influence the dynamics. 
With increasing obstacle density the system exhibits a gradual transition from delocalized to localized dynamics. Subdiffusion with constant exponents can be identified for up to three decades in time. 
However, our results show that the system exhibits fundamentally different dynamics from the LM. Whereas for the single-particle case a mapping onto the LM transition is possible \cite{Schnyder2015}, we find here that the universality of the dynamics breaks down. The interaction of particles with each other makes each particle's energy time-dependent; the free area available to it changes with each collision with other mobile particles.
In that sense, the free volume in soft systems is dynamic -- not static as in the LM -- with drastic consequences. Mobile particles help each other over potential barriers, speeding up the dynamics when the mobile particle density is increased. This is impossible for interacting hard disks, thus giving a rare example where \emph{soft and hard interactions are qualitatively different}. 
Even though exponents similar to the LM exponent may occur, we show that they merely represent \emph{effective, non-universal} exponents which are highly tunable via the particle interactions. Thus they should not be linked to the anomalous exponent of the LM. In experiments, where interactions are typically quite complex, the LM can therefore at best serve as a tool for qualitative interpretation.

\begin{figure}
\includegraphics{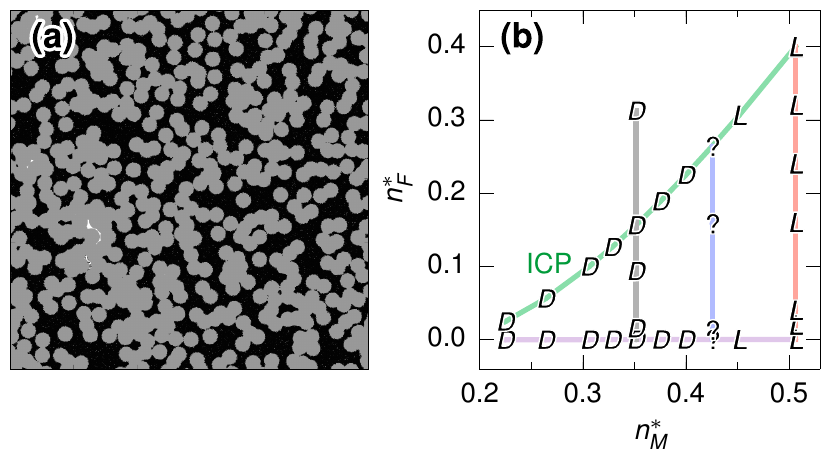}
\centering
\caption{\label{fig1} \emph{Simulation snapshots and state diagram.} (a)
All positions of the fluid particles over one simulation run at $n\reduced\matrx = 0.33$ and $n\fluid\reduced
= 0.127$ shown as black dots. Obstacles shown in grey. (b)
State diagram of the system with $D$ marking diffusive, $L$ marking
effectively localized states, and $?$ marking states where the dynamic
state was unclear on the time scale of the simulation. The
path crossing the critical point at finite $n\fluid\reduced$ is denoted
ICP (Interacting-particles Critical Path).}
\end{figure}
\paragraph*{Simulation details}
Matrix (index M) and fluid particles (F) interact via a smoothly
truncated, purely repulsive Weeks-Chandler-Andersen (WCA) potential \new{\cite{Weeks1971}, i.e. a Lennard-Jones potential which is truncated in the minimum and shifted.}
The size of fluid and matrix particles is given by
$\sigma\fluid$ and $\sigma\matrx$.  The matrix structures
are obtained as snapshots of equilibrated liquids.  For ensemble
averaging, we use 100 statistically independent matrix structures
 with up to $N\matrx = 16\,000$ particles each
at number density $n\matrx := N\matrx/L^2 = 0.278\,(\sigma\matrx)^{-2}$,
corresponding to system sizes of up to $L/\sigma\matrx = 240$. 
The energy coefficient for the interactions between matrix particles $\eps\matrx$ sets the simulation's energy unit. For more details, Ref.~\cite{Schnyder2015}.
Fluid particles are inserted into the frozen matrix with number
density $n\fluid:= N\fluid/L^2$, \cref{fig1}(a). 
Fluid particles interact with
matrix particles with coefficients $0.1 \eps\matrx$
and $(\sigma\matrx + \sigma\fluid)/2 =:
\sigma$.  For the interaction between fluid particles, we use $\eps\matrx$.  Newton's equations of motion are integrated
with the velocity-Verlet algorithm \cite{Binder2004} with time step
$7.2\cdot 10^{-4}t_0$ with $t_0 := [m(\sigma\matrx)^2/\eps\matrx]^{1/2}$
and $m=1.0$ the mass of a fluid particle.  
The fluid particles are
equilibrated using a simplified Andersen thermostat \cite{Andersen1980}
by randomly drawing their velocities from a Maxwell distribution every
100 steps for up to $1.4\cdot10^5t_0$.  
Up to 2400 fluid particles per host structure
are used for runs of up to nearly $10^6 t_0$. For the calculation of
time averages, 10 time origins per run are used, spaced equidistantly
over the whole simulation time. 

We use two control parameters: The interaction range between matrix and
fluid particles $\sigma:=(\sigma\fluid + \sigma\matrx)/2$ is tuned by
the \emph{diameter of the fluid particles} $\sigma\fluid$. This defines
the reduced number density $n\reduced\matrx:= n\matrx \sigma^2$ of the
matrix. \new{In addition, we vary the \emph{number density of fluid particles} $n\fluid$ by inserting more or fewer particles into the matrix.}
Both
$n\fluid$ and $\sigma\fluid$ change the reduced number density
$n\fluid^* := n\fluid\sigma\fluid^2$ of the fluid.  The control parameters
map out the state diagram of $n\fluid\reduced$ and $n\reduced\matrx$,
see \cref{fig1}(b).  To determine the dynamic state of the
systems, the mean-squared displacement (MSD) $\msd(t) = \av{|\vec r(t) -
\vec r(0)|^2}$ was calculated from the particle positions $\vec r(t)$ as a
time- and ensemble average.  The systems where the MSD became diffusive,
i.e.~$\msd(t) \sim 4Dt$ with diffusion coefficient $D$ for $t<7\cdot
10^5$, are marked delocalized, ``$D$''; states where the MSD converged
to a finite long-time limit are marked localized, ``$L$''. The remaining points are marked as
``$?$''.  
We discuss the dynamics along the marked paths.
The path along $n\fluid\reduced = 0$ represents the
ideal gas limit of non-interacting tracers, for which the rounding of
the transition was discussed recently~\cite{Schnyder2015}. Starting from
this confined ideal gas, $n\fluid\reduced$ was gradually increased for constant
$n\reduced\matrx$ to study the modification of the dynamics by the
interaction between fluid particles.
To study delocalized dynamics, the dynamics close to the localization
transition, and localized dynamics, respectively, we chose densities $n\reduced\matrx= 0.35$, $0.43$, and $0.51$.

\begin{figure}
\includegraphics{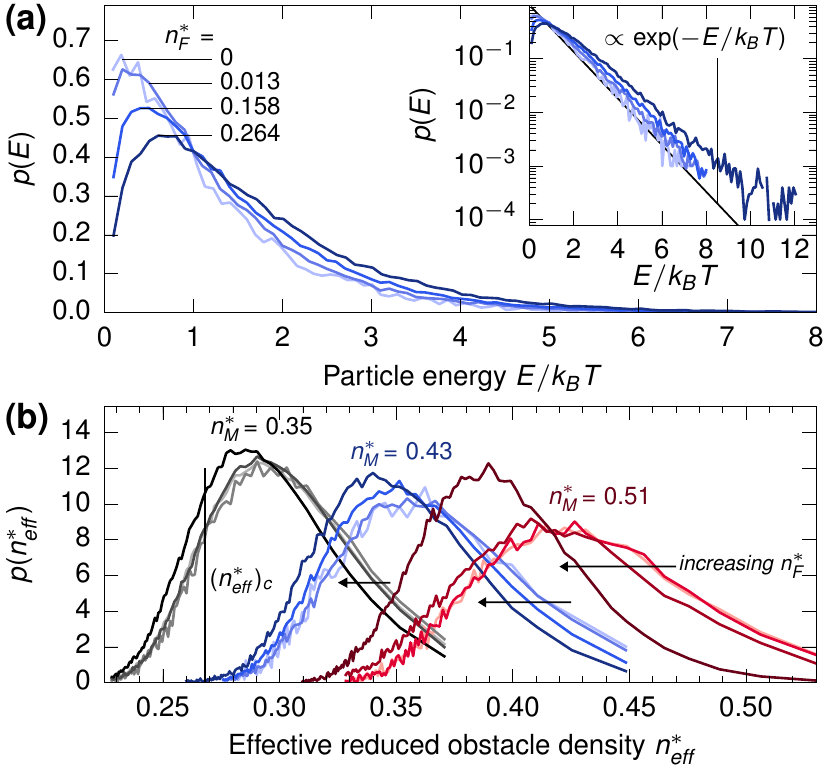}
\centering
\caption{\label{fig2} \emph{Energy and density distributions.} (a) Energy distributions $p(E)$ of
the fluid particles for $n\reduced\matrx= 0.43$ and a range of
$n\fluid\reduced$. Inset: same data in semilog. plot.
(b) Effective reduced obstacle density distribution $p(n\eff\reduced)$ calculated from $p(E)$ for $n\reduced\matrx= 0.35$, $0.43$ and $0.51$
and fluid densities $n\fluid\reduced = 0$, $0.016$, $0.094$, $0.313$,
$n\fluid\reduced = 0$, $0.013$, $0.158$, $0.264$, and $n\fluid\reduced
= 0$, $0.020$, $0.160$, $0.400$ respectively. Darker colors indicate
higher $n\fluid\reduced$. The critical effective density is $(n\eff\reduced)\crit
\approx 0.268$ \cite{Schnyder2015}.}
\end{figure}
\begin{figure*}
\includegraphics[scale = 1]{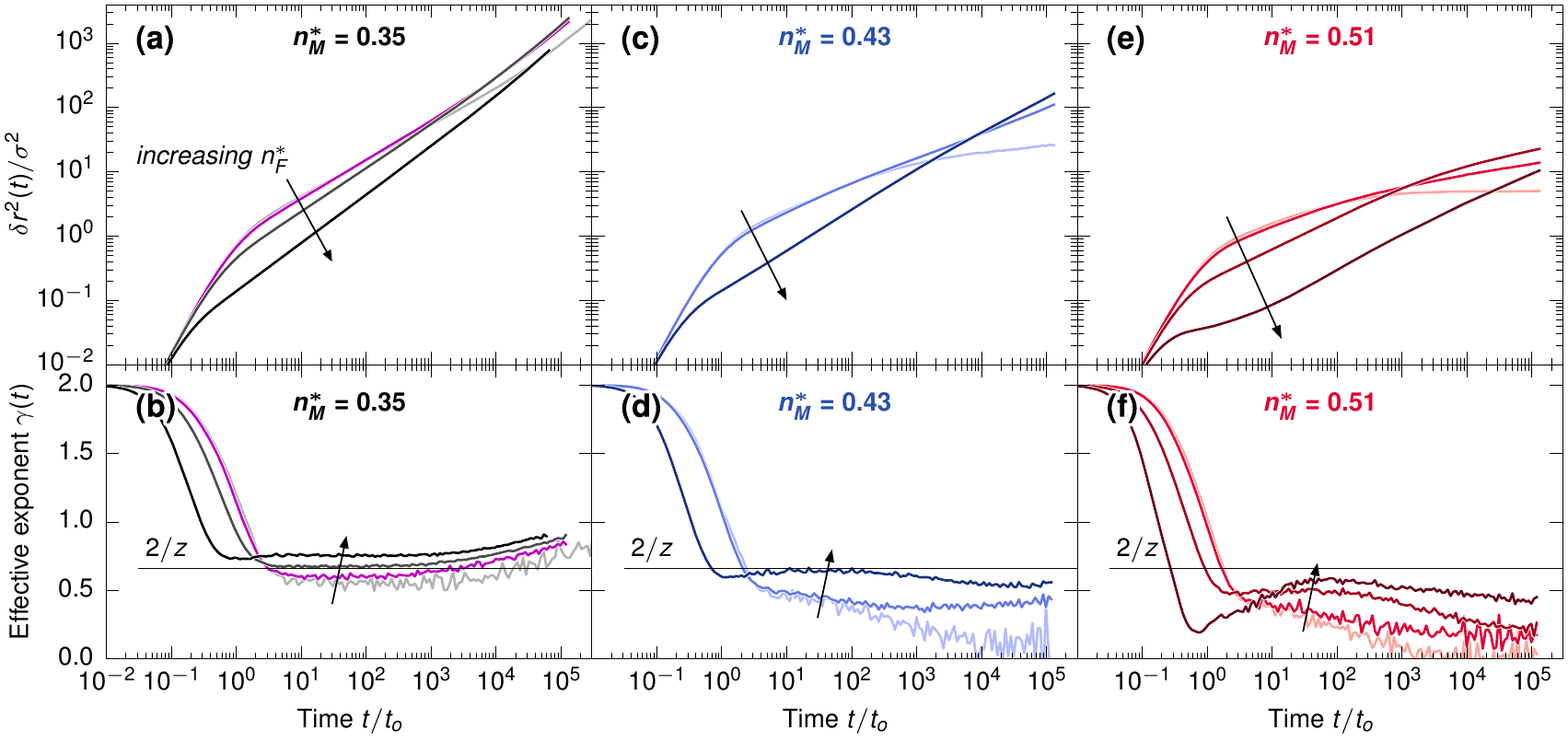}
\centering
\caption{\label{fig3} \emph{Anomalous transport.}
Mean-squared displacements $\msd(t)$ and
effective exponents $\gamma(t)$ for $n\reduced\matrx= 0.35$, $0.43$,
and $0.51$.
Fluid densities are  $n\fluid\reduced = 0$, $0.016$, $0.094$,
$0.313$ for (a, b) ($n\fluid = 0.016$ in purple for clarity);
$n\fluid\reduced = 0$, $0.013$, $0.264$ for (c, d); and $n\fluid\reduced = 0$,
$0.020$, $0.160$, $0.400$ for (e, f). In (b, d, f),
the critical exponent of the two-dimensional LM $2/z =
2/3.036$~\cite{Bauer2010} shown as black line.}
\end{figure*}

\paragraph*{Effective matrix density distributions}

If the energy $E$ of a tracer is conserved, an effective hard-disk
interaction diameter \new{$\sigma\eff(E,\sigma)$} can be calculated, mapping the 
system onto an effective LM with matrix density
\new{$n\eff\reduced := n\matrx (\sigma\eff)^2$} \cite{Schnyder2015}. But here the particles interact, exchange energy,
and have a time-dependent $\sigma\eff$.  While mapping a single tracer's dynamics onto the LM is impossible,
the time-independent probability \emph{distribution} $p(n\reduced\eff)$
for the whole system \emph{can} be calculated. With the energy of a
tracer $j$,
\begin{align*}
E_j = \frac{m \vec v_j^2}{2} 
     + \sum_{k\,\in\, M} V\MF(|\vec r_j - \vec r_k|)
     + \frac{1}{2} \sum_{\stackrel{k \neq j,}{k\,\in \,F}} 
       V\FF(|\vec r_j - \vec r_k|).
\end{align*}
the single-particle energy distribution $p(E)$ can be
calculated from the simulation as the histogram of tracer energy,
\cref{fig2}(a) for $n\reduced\matrx= 0.43$. The distributions have
a peak at small energies which decreases with increasing
$n\fluid\reduced$. The high-energy tail becomes more pronounced with
increasing $n\fluid\reduced$ but always decays exponentially, see inset.
The same holds for $n\reduced\matrx= 0.35$ and $0.51$ (not shown here).
From $p(E)$ the effective density distribution $p(n\eff\reduced)$
was calculated, see Ref.~\citenum{Schnyder2015} and
\cref{fig2}(b).  Large $E$ map onto small $\sigma\eff$.  In the
effective hard-disk system, the matrix ceases to percolate
at critical density $(n\eff\reduced)\crit$. $n\eff\reduced <
(n\eff\reduced)\crit$ correspond to delocalized and $n\eff\reduced >
(n\eff\reduced)\crit$ to localized states. The broad obstacle density
distributions are indicators of strongly averaged dynamics.

The distributions $p(n\reduced)$ for $n\reduced\matrx= 0.35$ are partly
on the delocalized side.  Increasing $n\fluid\reduced$ at constant
$n\reduced\matrx$ shifts the whole distribution towards
lower $n\eff\reduced$, indicating that on average more particles are
delocalized at any given time. This also shifts the localization transition towards higher $n\reduced\matrx$ at
constant $n\fluid\reduced$.  This generic shift of the distribution upon
increasing $n\fluid\reduced$ leads from localization to
delocalization in a system close to the localization transition, e.g.~at
$n\reduced\matrx= 0.43$: Whereas at small densities, $n\fluid\reduced
\leq 0.158$, the distribution $p(n\eff\reduced)$ is on the
localized side, at $n\fluid\reduced = 0.264$, delocalized states become
available. For $n\reduced\matrx= 0.51$ the distribution stays on
the localized side.

\paragraph*{Dynamics}
The MSD undergoes strong
changes on both sides of the transition as the fluid density
 is increased.  At $n\reduced\matrx= 0.35$ (\cref{fig3}(a)),
all investigated systems are delocalized, as anticipated from
$p(n\eff\reduced)$. The confined ideal gas ($n\fluid\reduced = 0$)
shows subdiffusion on intermediate times before becoming diffusive
at long times.  The MSD at $n\fluid\reduced = 0.016$ (purple), while
nearly unchanged at small and intermediate times, shows considerably
enhanced long-time diffusion. This happens even though the energy
distribution is nearly the same.

The speeding up of the dynamics stems from the \emph{exchange}
of energy between particles: When particles exchange energy, more
particles have a high energy at some point during the simulation
and can escape void pockets and explore more of the void
space.  This is only possible in systems with soft interactions, where
the barriers between void pockets are finite and thus surmountable.

The MSD for $n\fluid\reduced = 0.094$ is
suppressed on short and intermediate times compared to the confined
ideal gas because collisions of particles with their neighbors slow down
the exploration of the void volume. But at long times the MSD catches
up with the MSD at $n\fluid\reduced = 0.013$ and overtakes it.
At $n\fluid\reduced = 0.313$, the dynamics is further slowed down. At long times, the diffusion has slowed down
compared to the systems at intermediate $n\fluid\reduced$ but is still
 similar to the confined-ideal-gas case. This happens even though a
larger fraction of particles is delocalized at any given time than in
the less dense systems, as inferred from $p(n\reduced)$. This
indicates competition at long times between a
speeding up via energy exchange and a slowing
down via caging by neighbors.

The effective exponent of the MSD $\gamma(t) := \dd (\log \msd(t))/\dd
(\log t)$ allows identifying regimes where the MSD
follows a power-law. All systems show
constant $\gamma(t)$ over about 3 decades in time, \Cref{fig3}(b), with values ranging from below the LM critical exponent $2/z =
2/3.036$ \cite{Bauer2010} for $n\fluid = 0$ to above it for denser
systems. Therefore, $\gamma(t)$ can be readily tuned via the fluid density. 
For $n\fluid\reduced = 0.094$, $\gamma(t)$
nearly matches $2/z$ which is accidental as the system is \emph{clearly diffusive} at long times. Observing a $\gamma(t)$
close to the LM value is thus not
enough to determine that a system is near-critical.

For the localized system at $n\reduced\matrx= 0.51$ a similar modification
of the dynamics is found (\cref{fig3}(e)). At $n\fluid\reduced = 0$
the MSD converges to a finite long time limit, which is a measure of the
localization length.  Increasing the density to $n\fluid\reduced
= 0.020$ leaves the dynamics on short and intermediate times unchanged
but strongly increases the long time limit, due to particles
pushing each other out of void pockets (observed in 
trajectories). The increase of the long-time limit is evident even though
the MSD does not fully converge during the simulation. All
this occurs without a significant change in $p(n\eff\reduced)$.
Increasing $n\fluid\reduced$ further leads to a slowing down of the
dynamics on intermediate time scales and first to a speeding up and then
a slowing down at long times.

The system at $n\reduced\matrx= 0.43$ is an intermediate case, \cref{fig3}(c,d). 
At $n\fluid\reduced = 0$, the system is localized
since $\gamma(t)$ decays to near 0. The corresponding
distribution $p(n\eff\reduced)$ implies that all particles are localized,
with a few being very close to the transition. As a
result, the MSD grows slightly at all times.
At higher $n\fluid\reduced$ the localization length grows while
the intermediate-time dynamics slows down. As the MSDs are not
diffusive on the time scale of the simulation it is impossible to tell
whether the systems at finite $n\fluid\reduced$ are delocalized. Still,
diffusion of the MSD is anticipated from the upward bend in
the effective exponent $\gamma(t)$ at long times, \cref{fig3}(d), and
the shift of $p(n\eff\reduced)$ over the critical point, \cref{fig2}(b).

\begin{figure}
\includegraphics{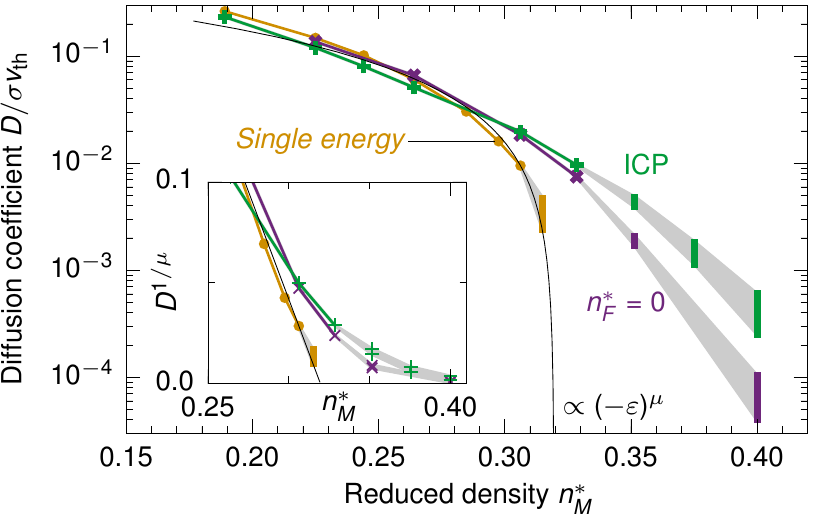}
\caption{\label{fig4} \emph{Rounding of the localisation transition.} Diffusion coefficient $D$ as function of $n\reduced\matrx$ for the single-energy case, for $n\fluid\reduced =
0$, and for $n\fluid = 0.625$ (ICP in \cref{fig1}).  
Connected symbols are directly obtained from
the MSD, while error bars are obtained in finite
size analysis, see ref.~\citenum{Schnyder2015}. The solid line $\propto
(-\eps)^\mu$ with $\eps = (n\reduced-n\reduced\crit) / n\reduced\crit$,
critical point $n\reduced\crit = 0.32$, and conductivity exponent
$\mu = 1.309$ of the LM, serves as guide to the eye.  Inset:
rectification plot of same data.}
\end{figure}

The scaling properties of the dynamics near the localization transition
were tested by crossing the transition along a path with constant
$n\fluid = 0.625$ while varying $n\reduced\matrx$.  Along this path,
$n\fluid\reduced$ is high enough so that the dynamics is considerably
modified by the interactions.  The diffusion coefficient $D$ is similar to that of the confined ideal gas ($n\fluid\reduced = 0$), and does not follow
the LM critical scaling, \cref{fig4}.  This is in sharp
contrast to the case where the tracers of the confined ideal gas are
all set to the same energy, for which said scaling has been identified \cite{Schnyder2015}.  The rounding of the transition is
even more clearly visible in the rectification plot in the inset of
\cref{fig4}, where data obeying the critical asymptote would fall on a
straight line.

\paragraph*{Discussion}

Similar to the hard-disk Lorentz model, the particles show
a localization transition as well as subdiffusion in the mean-squared displacement, extending up to 3 
decades in time. The associated effective exponents are tunable via the particle interactions
and may even match the Lorentz model value (cf.~the
finding in Ref.~\cite{Kim2009}). 
However, whereas interacting soft-disks can push each other out of void pockets over barriers in this soft version of the Lorentz model,
the situation is markedly different in the
corresponding hard-disk model
where the barriers, formed by closed pockets of hard obstacles, are always
infinite. In the hard-disk case, localization transitions similar to those
in the ``original'' Lorentz model with a single tracer particle can be
expected and they have in fact been observed in simulations
\cite{Kurzidim2009,Kurzidim2010,Kurzidim2011}.

The tune-ability of the exponents and the speeding up of the dynamics indicate the breaking of universality. Interestingly, similar speeding up of the dynamics was observed for a binary mixture of colloids \cite{Voigtmann2009}, where however varying the density of the more mobile component influences the structure of the less mobile component. Similarly, in a recent study of the MCT of fluids in random potential landscapes~\cite{Konincks2017}, a qualitatively similar speed-up was reported, as well. However, the theory still predicts a sharp localisation transition.

Our work demonstrates that the wide range of exponents seen, e.g. in crowding experiments of cellular fluids is \emph{most likely a result of the soft interactions between the components of those systems}. One expects therefore that crowding phenomena in cells cannot in general be associated with universal anomalous diffusion.

\acknowledgements
We acknowledge financial support by the DFG Research Unit FOR1394
``Nonlinear Response to Probe Vitrification'' (HO 2231/7-2, project P8).


\end{document}